\documentclass[%
reprint,
 amsmath,amssymb,
 aps,
pre,
]{revtex4-1}

\usepackage{dcolumn}
\usepackage{bm}
\usepackage{graphicx}
\usepackage{subfigure}
\usepackage{color}
\usepackage{epsfig}
\usepackage{amssymb}
\usepackage{latexsym}

\newtheorem{prop}{Proposition}[section]

\newtheorem{cor}{Corollary}

\newtheorem{lm}{Lemma}

\newtheorem{thm}{Theorem}

\newcommand{\bthm}{\begin{thm}}
\newcommand{\ethm}{\end{thm}}

\newcommand{\bcor}{\begin{cor}}
\newcommand{\ecor}{\end{cor}}
\newcommand{\bprop}{\begin{prop}}
\newcommand{\eprop}{\end{prop}}
\newcommand{\blm}{\begin{lm}}
\newcommand{\elm}{\end{lm}}
\newcommand{\beq}{\begin{equation}}
\newcommand{\eeq}{\end{equation}}
\newcommand{\ber}{\begin{eqnarray}}
\newcommand{\eer}{\end{eqnarray}}

\newenvironment{proof1}{\begin{trivlist}\item[]{\bf Proof:\hspace{2mm}}}{\hfill$\blackbox$\end{trivlist}}

\newcommand{\blackbox}{\vrule height7pt width5pt depth1pt}

\newcommand{\bit}{\begin{itemize}}
\newcommand{\eit}{\end{itemize}}
\newcommand{\ben}{\begin{enumerate}}
\newcommand{\een}{\end{enumerate}}
\newcommand{\bdesc}{\begin{description}}
\newcommand{\edesc}{\end{description}}
\newcommand{\beqarrn}{\begin{eqnarray*}}
\newcommand{\eeqarrn}{\end{eqnarray*}}
\newenvironment{proofof}[1]{\begin{trivlist}\item[]{\bf Proof of #1:\hspace{2mm}
}}{\hfill\blackbox\end{trivlist}}
\newcommand{\bproofof}{\begin{proofof}}
\newcommand{\eproofof}{\end{proofof}}
\newenvironment{rem}{\begin{trivlist}\item[]{\bf
Remark:}\hspace{4mm}}{\end{trivlist}}
\newcommand{\brem}{\begin{rem}}
\newcommand{\erem}{\end{rem}}
\newenvironment{rems}{\begin{trivlist}\item[]{\bf
Remarks}\begin{itemize}}{\end{itemize}\end{trivlist}}
\newcommand{\brems}{\begin{rems}}
\newcommand{\erems}{\end{rems}}
\newtheorem{fact}{Fact}
\newcommand{\bfact}{\begin{fact}}
\newcommand{\efact}{\end{fact}}
\newtheorem{examp}{Example}
\newcommand{\bexamp}{\begin{examp}\rm}
\newcommand{\eexamp}{\end{examp}}
\newtheorem{defn}{Definition}
\newcommand{\bdefn}{\begin{defn}\rm}
\newcommand{\edefn}{\end{defn}}

\newtheorem{prob}{Problem}
\newcommand{\bprob}{\begin{prob}}
\newcommand{\eprob}{\end{prob}}

\newcommand{\bvtm}{\begin{verbatim}}
\newcommand{\bfig}{\begin{figure}}
\newcommand{\efig}{\end{figure}}
\newcommand{\bcen}{\begin{center}}
\newcommand{\ecen}{\end{center}}

\long\def\comment#1{}

\def \n2{{N_0 \over 2}}

\newcommand{\bP}[1]{{\mathbb{P}}\left[{#1}\right]}

\newcommand{\1}[1]{{\bf 1}\left[#1\right]}

\def \h5{\hspace{0.5in}}

\begin{document}

\preprint{APS/123-QED}

\title{Analysis of complex contagions in random multiplex networks}

\author{Osman Ya\u{g}an}
\author{Virgil Gligor}%
\affiliation{%
ECE Department and CyLab, 
Carnegie Mellon University, Pittsburgh, PA 15213 USA}%


%

\date{\today}

\begin{abstract}
We study the diffusion of influence in random {\em multiplex} networks where links can be 
of $r$ different types, and for a given {\em content} (e.g., rumor, 
product, political view), each link type is associated with a content dependent
parameter $c_i$ in $[0,\infty]$ that measures the relative bias type-$i$ links
have in spreading this content.  In this setting, we propose a 
linear threshold model of contagion where nodes switch state if their
\lq\lq perceived" proportion of active {\em neighbors} exceeds a threshold $\tau$. Namely,
a node connected to $m_i$ active neighbors and $k_i-m_i$ inactive neighbors via type-$i$ links
will turn active if $\sum{c_i m_i}/\sum{c_i k_i}$ exceeds its threshold $\tau$. Under this model,
we obtain the condition, probability and expected size of {\em global} spreading events. 
Our results extend the existing work on complex contagions in several directions by (i) providing 
solutions for coupled random networks whose vertices are neither identical nor disjoint, 
(ii) highlighting the
effect of content on the dynamics of complex contagions, and 
(iii) showing that content-dependent
propagation over a multiplex network leads to
a {\em subtle} relation between the giant vulnerable component of the graph 
and the global cascade condition that is not seen in the existing models in the literature.

\begin{description}
\item[PACS numbers]{89.75.Hc, 87.23.Ge}

\end{description}
\end{abstract}
\maketitle

\section{Introduction}

In the past decade, there has been an increasing interest 
in studying dynamical processes on real-world complex networks. An interesting 
phenomenon that seems to occur in many such processes is the spreading of 
an initially localized effect throughout the whole (or, a very large part of the) network. 
These events are usually referred to as (information) {\em cascades} and
can be observed in processes as
diverse as adoption of cultural fads, the diffusion of belief, norms, and innovations in social networks 
\cite{WattsExternal,DoddsWatts,DuanChenLiuJin}, 
disease contagion in human and animal populations \cite{Murray, AndersonMay},
failures in {\em interdependent} infrastructures 
\cite{Buldyrev,BuldyrevShereCwilich, ChoGohKim,
CohenHavlin,
HuangGaoBuldyrevHavlinStanley,
ShaoBuldyrevHavlinStanley,Vespignani,YaganQianZhangCochranLong,YaganQianZhangCochran},
rise of collective action to join a riot \cite{Granovetter}, and global spread of computer viruses 
or worms on the Web \cite{NewmanForrestBalthrop,BalthropForrestNewmanWilliamson}.

The current paper focuses on a class of dynamic processes
usually known as binary decisions with externalities \cite{WattsExternal}.
In this model,
nodes can be in either one of the two states: {\em active} or {\em inactive}. 
Each node is initially given a random threshold $\tau$ in $(0,1]$ drawn independently
from a distribution $P_{\textrm{th}}(\tau)$. Then, starting from a small number of active vertices,
nodes update their states (synchronously) in discrete time steps. An inactive node 
with $m$ active neighbors and $k-m$ inactive neighbors 
will be activated only if the fraction $\frac{m}{k}$ exceeds $\tau$; 
 once active, a node can 
not be deactivated. This model is sometimes 
referred to as the Watts' threshold model. 
However,
it was motivated by the 
seminal work of Schelling \cite{Schelling} who employed a threshold model 
to gain insight into residential segregation. 
Granovetter \cite{Granovetter} also studied this 
model in a fully mixed population (i.e, one where each individual can affect any
other one regardless of network topology) to characterize riot behavior.   

The starting point of the current work  is the following basic observation.
Most existing studies on this subject are based on
the assumption that all the links in the network are of the same type; i.e., 
it is assumed that the underlying network is {\em simplex}.
However, in reality, links might be {\em classified} 
according to the nature of the relationship they represent, and
each link type may play a different role in different
cascade processes. For example, in the spread of a new consumer product amongst the population, 
a video game would be more likely to be promoted among high school classmates 
rather than among family members \cite{TangYuanMaoLiChenDai}; the situation 
would be exactly the opposite
in the case of a new cleaning product. 
Several other examples,
both intuitive {\em and} real-world, can be given
to show the relevance of 
link classification. A few of them include belief propagation in
a coupled social-physical network (links between distant Facebook friends vs. links
between close office-mates), 
cascading failures in {\em interdependent} networks (power links that are vulnerable to 
natural hazards vs. computer links that are vulnerable to viruses), and spread of worms or viruses over the Internet
(worms that spread via E-mail contacts vs. worms that exploit particular system vulnerabilities;
e.g., see the Internet (Morris) worm of November 2, 1988 \cite{Worm}).
 
With this motivation in mind, in this paper we study the cascade
processes in {\em multiplex} networks 
\cite{PadgettAnsell,SzellLambiotteTurner,LeeKimChoGohKim,BrummittLeeGoh}: 
Assume that the links in the network are classified into $r$ different types
$1, \ldots, r$. For a given content (a product, view, rumor, or a source of failure), 
consider positive scalars $c_1, \ldots, c_r$,
such that $c_i$ quantifies the relative bias a type-$i$ link has in 
spreading this particular content; i.e.,
the larger the constant $c_i$, the more likely it is for the 
content to spread via type-$i$ links. 
Then, we assume that an inactive node with threshold 
$\tau$ gets activated if 
\begin{equation}
\frac{c_1 m_1 + c_2 m_2 + \ldots +c_r m_r}{c_1 k_1 + c_2 k_2 + \ldots + c_r k_r} \geq \tau
\label{eq:perceived_proportion}
\end{equation}
where $m_i$ (resp. $k_i$) is the number of {\em active} neighbors (resp. number of neighbors)
that the node is connected via a type-$i$ link. In other words, instead of using the fraction ${m}/{k}$
of one's active neighbors, we use the content-dependent quantity ${\sum_{i=1}^{r} c_i m_i }/{\sum_{i=1}^{r}c_i k_i}$,
hereafter referred to as the {\em perceived} proportion of active neighbors
\footnote{The notion of perceived proportion of active neighbors was first suggested by Granovetter  
\cite[pp. 1429]{Granovetter} as a way of
taking the {\em social structure}  into account in the study of cascading processes. There, he 
suggested, as an example, that the influence of friends would be twice that of strangers in a
fully mixed population; in the formulation (\ref{eq:perceived_proportion}),
this amounts to setting (with $r=2$) $c_1=2$ and $c_2=1$, where links of type-$1$ are 
considered as friendship links, whereas links of type-$2$ are considered as links
with strangers.}. This formulation allows a more accurate
characterization of a node's influence on others' behavior with respect to 
spreading of various contents; the original case 
can easily be recovered by setting $c_1=\ldots=c_r=1$.
   
Under the condition (\ref{eq:perceived_proportion}) for adoption, we are 
interested in understanding whether
a single node can trigger a global cascade; i.e., whether
a linear fraction of nodes (in the
asymptotic limit) eventually becomes active when 
an arbitrary node is switched to the active state. 
For ease of exposition, we consider the case where
links are classified into {\em two} types; extension to $r$ types is straightforward. 
Assuming that each link type defines a sub-network which is constructed according to 
the configuration model \cite{newman2001random}, 
we find the conditions under which a global cascade is possible;
the precise definition of the model is given in Section \ref{sec:Model}.
In the cases where a global spreading event is possible, we find the exact probability 
of its taking place, as well as the final expected cascade size. 

These results 
constitute an extension 
of the results by Watts \cite{WattsExternal} in several 
directions: First, our work extends the previous results on single networks 
with arbitrary degree distribution 
to multiple overlay networks \cite{YaganQianInfoPropLong} where the 
vertex sets of the constituent networks are not disjoint (as in {\em modular}
networks \cite{Gleeson2008}). Second, by introducing the condition (\ref{eq:perceived_proportion})
for adoption, our model is capable of capturing 
the relative effect of {\em content} in the spread of influence, and our theory
highlights how different content may have different spreading characteristics
over the same network. Third, our analysis indicates that content-dependent
propagation over a network with classified links entails multiple notions of 
{\em vulnerability} (with respect to each link type), resulting in a 
{\em directed} subgraph on vulnerable nodes. This leads
to a {\em subtle} relation between the giant vulnerable component of the graph 
and the global cascade condition in a manner different than the
existing models \cite{WattsExternal, DoddsPayne, Gleeson2008,PayneHarrisDodds,MelnikWardGleesonPorter}.

Very recently, Brummitt et al. \cite{BrummittLeeGoh} also studied the dynamics of
cascades in multiplex networks, but under a different formulation then ours. There, they
assumed that a node becomes active if the fraction of its active neighbors in {\em any} link type 
exceeds a certain threshold. With the notation introduced so far, this condition amounts
to
\begin{equation}
\max_{i=1,\ldots, r}\left(\frac{m_i}{k_i}\right ) \geq \tau.
\label{eq:condition_Brummitt}
\end{equation}
In setting (\ref{eq:condition_Brummitt}), the authors studied 
the threshold and the size of global cascades 
and found that multiplex networks are more vulnerable to cascades as compared to
simplex networks. Although formulation (\ref{eq:condition_Brummitt}) might be relevant for 
certain cases, it can not capture the effect of content in the cascade process.
Furthermore, condition 
(\ref{eq:perceived_proportion}) proposed here enables  more general observations in terms
of the vulnerability of multiplex networks: Depending on the content 
parameters $c_1, \ldots, c_r$, a multiplex network can be
more, less, or equally vulnerable to cascades as compared to a simplex network with 
the same total degree distribution; e.g.,  see Section \ref{sec:Simu}. 
In fact, it always holds that
\[
\min_{i= 1,\ldots, r} \left(\frac{m_i}{k_i}\right) \leq \frac{\sum_{i=1}^{r} c_i m_i}{\sum_{i=1}^{r} c_i k_i}
\leq \max_{i= 1,\ldots, r} \left(\frac{m_i}{k_i}\right).
\]
However, it is worth noting that the results obtained here do not
contain those of \cite{BrummittLeeGoh} since one can not
select $c_1, \ldots, c_r$ such that ${\sum_{i=1}^{r} c_i m_i}/{\sum_{i=1}^{r} c_i k_i}
= \max_{i= 1,\ldots, r} \left({m_i}/{k_i}\right)$ holds for all possible $\{m_i, k_i\}_{i=1}^{r}$.
 
The paper is structured as follows. In Section \ref{sec:Model} we give the 
details of the system model. Analytical results 
regarding the condition, probability, and the size of
global cascades over the system model are given in Section \ref{sec:Results},
while in Section \ref{sec:Simu} we present numerical results 
that illustrate the main findings of the paper. We close with some 
remarks in Section \ref{sec:Conclusion}.

\section{Model Definitions}
\label{sec:Model}

For illustration purposes we give the model definitions
in the context of an overlay social-physical
network. We start with a set of individuals in the population represented
by nodes $1, \ldots, n$. Let 
$\mathbb{W}$ stand for the {\em physical} 
network of individuals that characterizes the
possible spread of influence through {\em
reciprocal} (i.e., mutual) communications; a link represents a
reciprocal communication if there is a message exchange in both
directions over the link. Examples of reciprocal communications include face-to-face communications,
phone calls, chats, or mutual communications through 
an online social networking website. 
Next, we let $\mathbb{F}$ stand for
a network that characterizes the spread of influence through non-reciprocal
communications in an online social networking web site, e.g., Facebook 
\footnote{
We remind that
these definitions are given merely for illustration purposes
and do not effect our technical results. Our intuition is to distinguish people with
{\em close} relationships (as understood from their engagement in two-way communications)  
and those that are merely Facebook {\em friends} who receive information and status
updates from one another but never talk to each other. Recent statistics show that \cite{Merlow},
on average, a user with $500$ friends in Facebook 
engages in a mutual communication with only $13$ of them; a number likely to represent
one's close relationships. Also, we refer to
the network $\mathbb{W}$ as a {\em physical} one since its links appear between people that have 
close relationships. For instance, we regard 
a mother using Facebook to communicate with her daughter (who lives abroad)
as if they belong to each other's physical network.}.
We assume that the physical network
$\mathbb{W}$ is defined on the vertices $\mathcal{N}=\{1,2,\ldots, n\}$ implying that
each individual in the population is a member of $\mathbb{W}$.
Considering the fact that not everyone in the population is a member of
online social networks, we assume that
the network $\mathbb{F}$ is defined on the vertex set $\mathcal{N}_F$
where
\begin{equation}
 \bP{i
\in \mathcal{N}_F} = \alpha, \qquad i=1,\ldots, n,
\label{eq:members_face}
\end{equation}
In other words, we assume that each node in $\mathcal{N}$ is a {\em member} of
$\mathbb{F}$ with probability $\alpha \in (0,1]$ independently from any
other node. 

We define the structure of the networks $\mathbb{W}$ and $\mathbb{F}$ 
through their respective degree distributions $\{p_k^w\}$ and $\{p_k^f\}$.
In other words, for each $k=0,1,\ldots$, a node in $\mathbb{W}$ (resp.  in $\mathbb{F}$) 
has degree $k$ with probability $p_k^w$ (resp. $p_k^f$).  
This corresponds to generating both networks (independently)
according to the {\em configuration} model
\cite{Bollobas,MolloyReed}. Then, we consider an overlay network
 $\mathbb{H}$
that is constructed by taking the {\em union} of $\mathbb{W}$ and
$\mathbb{F}$. In other words, for any distinct pair of nodes
$i,j$, we say that $i$ and $j$ are adjacent in the network
$\mathbb{H}$, denoted $i \sim_{\mathbb{H}} j$, as long as at least
one of the conditions \{$i \sim_{\mathbb{W}} j$\} or \{$i
\sim_{\mathbb{F}} j$\} holds.

The overlay network $\mathbb{H}=\mathbb{W} \cup \mathbb{F}$ 
constitutes an ensemble of the {\em colored} degree-driven random 
graphs studied by S\"{o}derberg \cite{Soderberg2,Soderberg4}. 
Let $\{1,2\}$ be the space of possible colors (or types) 
of edges in $\mathbb{H}$; specifically, we let the edges in Facebook
be of type $1$,
while the edges in the physical network are said to be of type $2$. 
The {\em colored} degree of a node $i$ is given
by an integer vector $\boldsymbol{k}^{i}=(k^i_{1}, k^i_2) $, where 
$k_1^{i}$ (resp. $k^i_2$) stands for the number of Facebook edges 
(resp. physical connections) that are incident on node $i$. Clearly,
the plain degree of a node is given by $k=k_1+k_2$.
Under the given assumptions on the 
degree distributions of $\mathbb{W}$ and $\mathbb{F}$, the colored
degrees (i.e., $\boldsymbol{k}^1, \ldots, \boldsymbol{k}^n$)
will be independent and identically distributed according to a colored degree distribution
$\{p_{\boldsymbol{k}}\}$ such that 
\begin{equation}
p_{\boldsymbol{k}} = \left(\alpha p_{k_1}^f + (1-\alpha) \1{k_1=0}\right) \cdot
p_{k_2}^w,  \quad \boldsymbol{k}=(k_1,k_2)
\label{eq:colored_dist}
\end{equation}
due to independence of $\mathbb{F}$ and $\mathbb{W}$. The term
$(1-\alpha) \1{k_1=0}$ accommodates the possibility that a 
node is not a member of the online social network, in which case the number $k_1$ of
$\mathbb{F}$-edges is automatically zero.

Given that the colored
degrees are picked such that $\sum_{i=1}^{n}k_1^{i}$ and 
$\sum_{i=1}^{n}k_2^{i}$ are even, we construct $\mathbb{H}$ as in
\cite{Soderberg2,Soderberg4,newman2002spread}: 
Each node $i=1,\ldots,n$ is first
given the appropriate number $k_1^{i}$ and $k_2^i$ of stubs of type $1$
and type $2$, respectively. Then, pairs of these stubs that are of the
same type are picked randomly and connected together to form
complete edges; clearly, two stubs can be
connected together {\em only} if they are of the same type.
Pairing of stubs continues until none is left. 

Now, consider a complex contagion process in the 
random network $\mathbb{H}$. As stated in the Introduction, 
we let each node $i$ be assigned a binary 
value $\sigma(i)$ specifying its current state, {\em active} 
($\sigma(i)=1$) or {\em inactive} ($\sigma(i)=0$). 
Each node is initially given a random threshold $\tau$ in $(0,1]$ drawn independently
from a distribution $P_{\textrm{th}}(\tau)$. 
Nodes update their states synchronously at times $t=0,1,\ldots$.
An inactive node will be
activated at time $t$ if, at time $t-1$, its perceived proportion
of active neighbors exceeds {\em its} threshold $\tau$. 
Namely, for a given
content, let $c_1$, $c_2$ be positive
scalars that model the relative importance of type-$1$ and type-$2$
links, respectively, in spreading this content. 
Then, with
$\boldsymbol{k}=(k_1,k_2)$ denoting its colored degree, and 
$\boldsymbol{m}=(m_1,m_2)$ denoting its
number of active neighbors connected through a type-$1$ and type-$2$ link at
time $t-1$, 
respectively, a node will become active (at time $t$) with probability
\[
\bP{\frac{c_1  m_1+ c_2 m_2}{c_1  k_1+ c_2 k_2} \geq \tau}
:=F(\boldsymbol{m},\boldsymbol{k}).
\]
Hereafter, $F(\boldsymbol{m},\boldsymbol{k})$
will be referred to as the {\em neighborhood influence response function} 
\cite{DoddsWatts,HackettMelnikGleeson}. 
To simplify the notation a bit, 
we let $c:= {c_1}/{c_2}$  for $c_1,c_2 > 0$ so that we have
\begin{equation}
F(\boldsymbol{m},\boldsymbol{k}) =\bP{\frac{c  m_1+  m_2}{c  k_1+ k_2} \geq \tau}.
\label{eq:response_function}
\end{equation}

The effect of content on the response of nodes can easily be inferred
from (\ref{eq:response_function}):
For instance, $c<1$ (resp. $c>1$) means that the 
current content is more likely to be promoted through type-$2$ edges (resp. type-$1$ edges).
The special case $c=1$ corresponds to the situations where both types 
of links have equal effect in spreading the content and  the
response function (\ref{eq:response_function}) reduces to that 
of a standard threshold model \cite{WattsExternal}. 
In the limit $c \to 0$ (resp. $c \to \infty$), we see that type-$1$ (resp. type-$2$) 
edges
have no effect in spreading the content and the network $\mathbb{H}$
becomes identical to a single network $\mathbb{W}$ (resp. $\mathbb{F}$) 
for the purposes
of the spread of this particular content.

\section{Analytic Results}
\label{sec:Results}

\subsection{Condition and Probability of Global Cascades}
\label{sec:Results_cond_and_prob}

We start our analysis by deriving the condition and probability of global spreading events in
overlay social-physical network $\mathbb{H}$.
In most existing works \cite{WattsExternal, DoddsPayne, Gleeson2008,PayneHarrisDodds}, 
the possibility of a global spreading event 
hinges heavily on the existence of a {\em percolating} cluster of nodes
whose state can be changed by only one active neighbor;
these nodes are usually referred to as {\em vulnerable}.
In other words, the condition for a global cascade to take place 
was shown to be equivalent to the existence of a {\em giant} 
vulnerable cluster in the network;
i.e., fractional size of the largest vulnerable cluster
being bounded away from 
zero in the asymptotic limit $n \to \infty$. The probability of triggering
a global cascade was then shown to be equal to the fractional size of the {\em extended}
vulnerable cluster, which contains nodes that have links to at least one node
in the giant vulnerable component.

Here, we will show that the situation is more complicated
unless the content parameter $c$ is unity. The subtlety arises
from the need for defining the notion of vulnerability
with respect to (w.r.t.) {\em two} different neighborhood relationships. Namely,
a node is said to be $\mathbb{W}$-vulnerable (resp. $\mathbb{F}$-vulnerable), 
if its state can be changed by a single link in 
$\mathbb{W}$ (resp. in $\mathbb{F}$) that connects it to an active node;
a node is simply said to be vulnerable, if it is vulnerable w.r.t. at least one of the networks.
Note that unless $c=1$, a node can be $\mathbb{F}$-vulnerable 
but not $\mathbb{W}$-vulnerable,
or vice versa. Therefore, an active vulnerable node does not necessarily activate 
all of its vulnerable neighbors, and the ordinary definition of a vulnerable component 
becomes vague. Here, we choose a natural definition of a vulnerable component in
the following manner: {\it A set of nodes that are vulnerable w.r.t. at least one 
of the networks are said to form a vulnerable component if in the subgraph
induced by this set of nodes,
activating any node
leads to the activation of all the nodes in the set.} 

In fact, the above definition of a vulnerable component
coincides with that of a {\em strongly connected component} 
\cite{DorogovtsevMendesSamukhin,newman2001random} in a {\em directed}
graph. To see this, consider the subgraph of vulnerable nodes in $\mathbb{H}$. This subgraph
forms a directed network, where a (potentially
bi-directional) $\mathbb{F}$-link
between nodes $i$ and $j$ will have the direction from $i$ to $j$ (resp. $j$ to $i$)
only if $j$ (resp. $i$) is $\mathbb{F}$-vulnerable; similar definitions determine the directions
of $\mathbb{W}$-links. There exist several definitions for the components of a directed graph, but we 
use that given by Bogu\~{n}\'a and Serrano \cite{BogunaSerrano} which is
adopted from \cite{DorogovtsevMendesSamukhin}. Namely, for a given vertex, 
its {\em out-component} is defined as the set of vertices that are reachable from it. Similarly,
the {\em in-component} of a vertex is the set of nodes that can reach that vertex.
Then, the giant out-component (GOUT) of a graph is defined as the set of nodes with {\em infinite}
in-component, whereas the set of nodes that have infinite out-component defines the giant
in-component (GIN). Finally, the giant strongly-connected component (GSCC) of the graph is given by the intersection
of GIN and GOUT. By definition, any pair of nodes in the GSCC are connected to each other
via a {\em directed} path.

The picture is now clear. According to the definition adopted here,  
the giant vulnerable component of the network corresponds to the GSCC of the
subgraph induced by vulnerable nodes. Moreover, global cascades 
can take place if there exists a linear fraction of vulnerable 
nodes whose out-component is infinite; i.e., the global cascade 
condition corresponds to the appearance of GIN
amongst the vulnerable nodes of $\mathbb{H}$.
Finally, the probability of triggering a global cascade will be given by the
fractional size of the {\em extended} GIN (EGIN), that contains GIN {\em and} vertices that are not vulnerable
but, once activated, can activate a node in GIN. 

In principle, it is possible for a directed network to have GIN but no GSCC \footnote{Consider a
network on vertices $\{1, \ldots, n\}$ with edges 
$1 \to 2 \to 3 \to \cdots \to n-1 \to n$ where $i \to j$ refers to an edge directed from 
$i$ to $j$. In the limit $n \to \infty$, 
a positive fraction of nodes have
{\em infinite} in- and out-components, but the network has no strongly connected component
since for each node, its in-component and out-component are disjoint.}, raising the possibility of
observing global cascades even 
when there is {\em no} giant vulnerable cluster in the network; this possibility would contradict
the previous results \cite{WattsExternal, DoddsPayne, Gleeson2008,PayneHarrisDodds}.
However, in all models
that appeared in the literature to date \cite{DorogovtsevMendesSamukhin,BogunaSerrano,Dousse},
it was shown that GIN, GOUT and GSCC appear simultaneously in the network.
In our case, since the condition and probability of global cascades can be obtained
by analyzing only GIN (and EGIN), we do not give an analysis to show
the simultaneous appearance of GIN and GSCC; instead, this step is taken via simulations
in Section \ref{sec:Simu}. From here onwards,
GSCC, GOUT and GIN refer to respective components 
of the {\em vulnerable nodes} in $\mathbb{H}$ even if it is not said so explicitly.

We now turn to computing
the probability (and condition) of triggering a global cascade by 
finding the size of EGIN of vulnerable nodes in the network $\mathbb{H}$.
This will be done by
considering a branching process which starts by activating an arbitrary node,
and then recursively reveals  the largest number of vulnerable nodes 
that are reached {\em and}
activated by exploring its neighbors. 
Utilizing the standard approach on
generating functions \cite{newman2001random,newman2002spread},
we can then determine the condition
for the existence of GIN as well as fractional size of EGIN; note that by definition
EGIN exists if and only if GIN does.
This approach is valid long as
the initial stages of the branching process is locally 
tree-like, which holds in this case as the clustering 
coefficient of colored degree-driven networks scales like $1/n$ as n gets
large \cite{Soderberg3}.

Throughout, we use
$\rho_{\boldsymbol{k},1}$ (resp. $\rho_{\boldsymbol{k},2}$) 
to denote the probability that a node is $\mathbb{F}$-vulnerable 
(resp. $\mathbb{W}$-vulnerable). 
In other words, $\rho_{\boldsymbol{k},1}$ (resp. $\rho_{\boldsymbol{k},2}$)
is the probability that an inactive node with 
colored degree $\boldsymbol{k}$ becomes active when it has
only one active neighbor in $\mathbb{F}$ (resp. in $\mathbb{W}$) 
and zero active neighbor in $\mathbb{W}$ (resp. in $\mathbb{F}$).
We also use
$\rho_{\boldsymbol{k},1\cap2}$ to denote the probability that
a node with colored degree $\boldsymbol{k}$ 
is both $\mathbb{F}$-vulnerable and $\mathbb{W}$-vulnerable.
In the same manner, we use $\rho_{\boldsymbol{k},1\cap2^c}$, 
$\rho_{\boldsymbol{k},1^c\cap2}$, and $\rho_{\boldsymbol{k},1^c\cap2^c}$,
to denote the probabilities that a node is $\mathbb{F}$-vulnerable but
not $\mathbb{W}$-vulnerable, $\mathbb{W}$-vulnerable but not $\mathbb{F}$-vulnerable,
and neither $\mathbb{F}$-vulnerable nor $\mathbb{W}$-vulnerable, respectively.
More precisely, we set
 \begin{eqnarray}
 \rho_{\boldsymbol{k},1\cap2} &=& \bP{\frac{c}{c k_1+k_2} \geq \tau ~~~\mbox{and}~~~ 
 \frac{1}{c k_1+k_2} \geq \tau}
\nonumber \\
  \rho_{\boldsymbol{k},1^c \cap2} &=& \bP{\frac{c}{c  k_1+k_2} < \tau ~~~\mbox{and}~~~ 
 \frac{1}{c  k_1+k_2} \geq \tau}.
 \nonumber
 \end{eqnarray}
 Similar relations define $\rho_{\boldsymbol{k},1}$, 
 $\rho_{\boldsymbol{k},2}$
 $\rho_{\boldsymbol{k},1^c\cap2}$, and 
 $\rho_{\boldsymbol{k},1^c\cap2^c}$. It is clear that
 if $c=1$, $\rho_{\boldsymbol{k},1}=\rho_{\boldsymbol{k},2}=\rho_{\boldsymbol{k},1\cap2}$,
 whereas $\rho_{\boldsymbol{k},1^c \cap2} = \rho_{\boldsymbol{k},1 \cap2^c} =0$.

We now solve for the survival probability of the aforementioned
branching process by using the mean-field approach based on the
generating functions \cite{newman2001random,newman2002spread}. 
Let $g_1(x)$ (resp. $g_2(x)$) denote the
generating functions for the {\em finite} number of nodes reached by following a 
type-$1$ (resp. type-$2$) edge in the above branching process. The generating functions 
$g_1(x)$ and $g_2(x)$ satisfy the self-consistency equations
\begin{eqnarray}
g_1(x)&=&x \sum_{\boldsymbol{k}} \frac{k_1 p_{\boldsymbol{k}}}{<k_1>}
\cdot \rho_{\boldsymbol{k},1} \cdot g_1(x)^{k_1-1} g_2(x)^{k_2} 
\label{eq:g_1}
\\ \nonumber
&&+ x^{0} \sum_{\boldsymbol{k}} \frac{k_1 p_{\boldsymbol{k}}}{<k_1>}
(1-\rho_{\boldsymbol{k},1})
\\
g_2(x)&=& x \sum_{\boldsymbol{k}} \frac{k_2 p_{\boldsymbol{k}}}{<k_2>}
\cdot \rho_{\boldsymbol{k},2} \cdot g_1(x)^{k_1} g_2(x)^{k_2-1} 
\label{eq:g_2}
\\ \nonumber 
& & \qquad + x^{0} \sum_{\boldsymbol{k}} \frac{k_2 p_{\boldsymbol{k}}}{<k_2>}
(1-\rho_{\boldsymbol{k},2}).
\end{eqnarray}

The validity of (\ref{eq:g_1}) can be seen as follows:
The explicit factor $x$ accounts for the initial vertex that is arrived at. The factor 
$k_1p_{\boldsymbol{k}}/<k_1> $ gives the {\em normalized} probability that an edge of 
type $1$ is attached (at the other end) to a vertex
with colored degree $\boldsymbol{k}$. Since the arrived node is reached 
by a type-$1$ link, it needs to be $\mathbb{F}$-vulnerable
to be added to the vulnerable component. 
If the arrived node is indeed $\mathbb{F}$-vulnerable
(happens with probability $\rho_{\boldsymbol{k},1}$)
it can activate other nodes via its remaining $k_1-1$
edges of type-$1$ and $k_2$ edges of type-$2$. 
Since the number of vulnerable nodes reached by each of its
type-$1$ (resp. type-$2$) links is 
generated in turn by $g_1(x)$ 
(resp. $g_2(x)$) we obtain the 
term $g_1(x)^{k_1-1} g_2(x)^{k_2}$ 
by the powers property of generating functions
\cite{newman2001random,newman2002spread}. Averaging
over all possible colored degrees $\boldsymbol{k}$ gives the 
first term in (\ref{eq:g_1}).
The second term with the factor $x^{0}=1$ accounts for the 
possibility that the arrived node is {\em not} $\mathbb{F}$-vulnerable and 
thus is not included in the cluster.
The relation (\ref{eq:g_2}) can be validated via 
similar arguments.

Using the relations (\ref{eq:g_1})-(\ref{eq:g_2}), we now find the
{\em finite} number of vulnerable nodes reached and activated 
by the above branching process.
With $G(x)$ denoting the corresponding generating function, we get
\begin{equation}
G(x)=x \sum_{\boldsymbol{k}} p_{\boldsymbol{k}} g_1(x)^{k_1} g_2(x)^{k_2}.
\label{eq:g_e(x)}
\end{equation}
Similar to (\ref{eq:g_1})-(\ref{eq:g_2}), the relation (\ref{eq:g_e(x)})
can be seen as follows: The factor $x$ corresponds to the 
initial node that is selected arbitrarily and made active. The selected node has
colored degree $\boldsymbol{k}=(k_1,k_2)$ with probability $p_{\boldsymbol{k}}$.
The number of vulnerable nodes that are reached by each of its $k_1$ (resp. $k_2$)
branches of type $1$ (resp. type $2$)
is generated by $g_1(x)$ (resp. $g_2(x)$). 
This yields the term $g_1(x)^{k_1} g_2(x)^{k_2}$
and averaging over all possible colored degrees,
we get (\ref{eq:g_e(x)}).

We are interested in the solution of the recursive
relations (\ref{eq:g_1})-(\ref{eq:g_2}) for the case $x=1$. This case exhibits a
trivial fixed point $g_1(1) = g_2(1)=1$ which
yields $G(1)=1$ meaning that the underlying branching process is
in the subcritical regime and that {\em all} components have
finite size as understood from the conservation of probability.
However, the fixed point $g_1(1) = g_2(1)=1$ corresponds to the physical 
solution only if it is an {\em
attractor}; i.e., a stable solution to the recursion (\ref{eq:g_1})-(\ref{eq:g_2}).
The stability of this fixed point can be checked via linearization
of (\ref{eq:g_1})-(\ref{eq:g_2}) around $g_1(1) = g_2(1)=1$, which 
yields the Jacobian $\boldsymbol{J}_p$ given by
\begin{equation}
\boldsymbol{J}_p = 
\left[
\begin{array}{cc}
\frac{< (k_1^2 - k_1) \rho_{\boldsymbol{k},1}>}{<k_1>} & 
\frac{< k_1 k_2 \rho_{\boldsymbol{k},1}>}{<k_1>} \\
& \\
\frac{< k_1 k_2 \rho_{\boldsymbol{k}, 2}>}{<k_2>} &
\frac{< (k_2^2 - k_2) \rho_{\boldsymbol{k},2}>}{<k_2>}
\end{array}
\right].
\label{eq:J_p}
\end{equation}

If all the eigenvalues of $\boldsymbol{J}_p$ are less than one in
absolute value (i.e., if the spectral radius $\sigma(\boldsymbol{J}_p)$ of $\boldsymbol{J}$
is less than or equal to unity),
then the solution $g_1(1)=g_2(1)=1$ is stable and $G(1)=1$ becomes the
physical solution, meaning that with high probability 
GIN does not exist.
In that case, a global 
spreading event is not possible and the fraction $S$
of influenced individuals always tends to zero. However, if the 
spectral radius of $\boldsymbol{J}_p$ is 
larger
than unity, then another solution with $g_1(1),g_2(1)<1$
becomes the attractor of (\ref{eq:g_1})-(\ref{eq:g_2}) yielding
a solution with $G(1)<1$. In that case, global cascades are possible meaning
that
switching the state of an arbitrary node gives rise to a
global spreading event with {\em positive} probability, $P_{trig}$. 
In fact, the deficit $1-G(1)$ 
corresponds to the probability that an arbitrary node, once activated, 
activates an infinite number of
vulnerable nodes, which in turn corresponds to the probability of 
triggering a global cascade; i.e., we have 
\[
P_{trig}=1-G(1).
\]

We close this section by noting that $G(x)$ 
corresponds to the size of the extended component EGIN, not GIN; i.e., 
$1-G(1)$ gives the asymptotic size of EGIN as a fraction 
of the number of nodes $n$.  This is because, in (\ref{eq:g_e(x)})
we have ignored the possibility of the initial node being {\em not} vulnerable; this makes
sense since, in the cascade process, the initially selected node is {\em forced} to be active regardless of its
state of vulnerability. In order to obtain the size of GIN, one
should consider another generating function $H(x)$ that is given by multiplying (\ref{eq:g_e(x)}) with the 
probability  ($1- \rho_{\boldsymbol{k},1^c\cap2^c}$) that the initial node is vulnerable
and adding the term $x^{0} \sum_{\boldsymbol{k}} p_{\boldsymbol{k}}  \rho_{\boldsymbol{k},1^c\cap2^c}$.
The asymptotic size of GIN (as a fraction of $n$) 
would then be given by $1-H(1)$.

\subsection{Expected Cascade Size}

We now compute 
expected final  
size of a global cascade when it occurs. Namely, we will 
derive the asymptotic fraction of individuals that eventually 
become active when an
arbitrary node is switched to active state. Our analysis is based on the 
work by Gleeson and Cahalane \cite{GleesonCahalane} and 
Gleeson \cite{Gleeson2008} who derived the expected final size of
global spreading events on a wide range of networks. Their approach, which
is built on the tools developed for analyzing the zero-temperature 
random-field Ising model on Bethe lattices \cite{Sethna1993}, is
also adopted by several other authors; e.g., see 
\cite{BrummittLeeGoh,DoddsPayne,HackettMelnikGleeson,PayneHarrisDodds}.
 
The discussion starts with the following basic observation: If the network 
structure is locally tree-like (which holds here as noted before \cite{Soderberg3}), 
then we can replace $\mathbb{H}$ by a tree structure where at 
the top level, there is a single node say with colored degree 
$\boldsymbol{k}=(k_1,k_2)$.  In other words, the top node is connected to 
$k_1$ nodes
via Facebook links and $k_2$ nodes via physical links at the next 
lower level of the tree. Each of these $k_1$ (resp. $k_2$)
nodes have degree $\boldsymbol{k}'=(k_1',k_2')$  
with probability $\frac{k_1' p_{\boldsymbol{k}}}{<k_1>}$ (resp.
with probability $\frac{k_2' p_{\boldsymbol{k}}}{<k_2>}$), and they
are in turn connected to $k_1'-1$ (resp. $k_1'$)
nodes via Facebook links and $k_2'$ (resp. $k_2'-1$) 
nodes via physical links at the next lower level 
of the tree; the minus one terms are due to the links that 
connect the nodes to their parent at the upper level.

In the manner outlined above, we label the levels of the tree from $\ell =0$
at the bottom to $\ell \to \infty$ at the top of the tree. Without loss
of generality, we assume that nodes 
update their states starting from the bottom of the tree and proceeding towards 
the top. In other words, we assume that a 
node at level $\ell$
updates its state only after all nodes at the lower levels $0,1,\ldots, \ell-1$ 
finish updating. Now, define $q_{1,\ell}$ 
(resp. $q_{2,\ell}$) as the probability that a node at level $\ell$ of the tree,
which is connected to its unique parent by a type-$1$ link (resp. a type-$2$ link), is active
given that its parent at level $\ell+1$ is inactive. Then, consider a node at 
level $\ell+1$ that is connected to its parent at level $\ell+2$ by a type-$1$ link. This node 
has degree $\boldsymbol{k}=(k_1,k_2)$ with probability $\frac{k_1 p_{\boldsymbol{k}}}{<k_1>}$
and the probability that $i$ of its Facebook neighbors and $j$ 
of its physical network connections
are active is given by
\begin{equation}
 {{k_1-1}\choose{i}} q_{1,\ell}^i 
(1-q_{1,\ell})^{k_1-1-i} 
 {{k_2}\choose{j}} q_{2,\ell}^{j} (1-q_{2,\ell})^{k_2-j} .
 \label{eq:prob_of_active}
\end{equation}
The minus one term on $k_1$ accommodates the fact that 
the parent of the node under consideration is inactive by the assumption that
nodes update their states only after all the nodes at the lower levels finish 
updating. Further, 
the probability that a node becomes active when $i$ of its $k_1$ Facebook
connections, and $j$ of its $k_2$ physical connections are active is 
given by
\[
F((i,j),\boldsymbol{k}), \qquad \boldsymbol{k}=(k_1,k_2)
\]
by the definition of neighborhood response function 
$F(\boldsymbol{m},\boldsymbol{k})$.

Arguments similar to the above one leads to analogous relations 
for nodes that are connected to their unique parents by a type-$2$ link.
Combining these, and averaging over all possible degrees and 
all possible active neighbor combinations, we arrive at
the recursive relations 
\begin{eqnarray}
q_{1,\ell+1} = \sum_{\boldsymbol{k}} \frac{k_1 p_{\boldsymbol{k}}}{<k_1>}
\sum_{i=0}^{k_1-1}\sum_{j=0}^{k_2} F((i,j),\boldsymbol{k}) {{k_1-1}\choose{i}} q_{1,\ell}^i 
\hspace{.2cm}
\nonumber \\
\times (1-q_{1,\ell})^{k_1-1-i} 
 {{k_2}\choose{j}} q_{2,\ell}^{j} (1-q_{2,\ell})^{k_2-j} \hspace{.5cm}
 \label{eq:q_1}
 \\
 q_{2,\ell+1} = \sum_{\boldsymbol{k}} \frac{k_2 p_{\boldsymbol{k}}}{<k_2>}
\sum_{i=0}^{k_1}\sum_{j=0}^{k_2-1} F((i,j),\boldsymbol{k}) {{k_1}\choose{i}} q_{1,\ell}^i 
\hspace{.7cm}
 \label{eq:q_2}
 \\
  \times
(1-q_{1,\ell})^{k_1-i} 
 {{k_2-1}\choose{j}} q_{2,\ell}^{j} (1-q_{2,\ell})^{k_2-1-j}, \hspace{-.2cm}
\nonumber
\end{eqnarray}
for each
$\ell=0,1,\ldots$.
Under the assumption that nodes do not become
inactive once they turn active, the quantities $q_{1,\ell}$ and
$q_{2,\ell}$ are non-decreasing in $\ell$ and thus, they converge to a 
limit $q_{1,\infty}$ and $q_{2,\infty}$.
Further, the final fraction of active individuals $S$ is equal (in expected value)
to the probability that the node at the top of the tree 
becomes active. Thus, we conclude that
\begin{eqnarray}
S &=& \sum_{\boldsymbol{k}} p_{\boldsymbol{k}}
\sum_{i=0}^{k_1}\sum_{j=0}^{k_2} F((i,j),\boldsymbol{k}) {{k_1}\choose{i}} q_{1,\infty}^i 
(1-q_{1,\infty})^{k_1-i} 
\nonumber \\  
& & \qquad \qquad \qquad \quad \times {{k_2}\choose{j}} q_{2,\infty}^{j} (1-q_{2,\infty})^{k_2-j}.
\label{eq:S}
\end{eqnarray}

Under the natural condition $F((0,0),\boldsymbol{k})=0$, $q_{1,\infty}=q_{2,\infty}=0$
is the trivial fixed point of the recursive equations (\ref{eq:q_1})-(\ref{eq:q_2}). 
In view of (\ref{eq:S}), this trivial solution yields $S=0$ pointing out
the {\em non-existence} of global spreading events.
However, the trivial fixed point may not be stable and another solution
with $q_{1,\infty},q_{2,\infty}>0$ may exist. In fact, the condition for the
existence of a non-trivial solution can be obtained by checking the 
stability of the trivial fixed point via linearization at $q_{1,\ell}=q_{2,\ell}=0$. 
The entries of the corresponding Jacobian matrix $\boldsymbol{J}_s$ 
is given by
\[
\boldsymbol{{J}}_s
=\left[
\begin{array}{cc}
\frac{< (k_1^2 - k_1) \rho_{\boldsymbol{k},1}>}{<k_1>} & 
\frac{< k_1 k_2 \rho_{\boldsymbol{k}, 2}>}{<k_1>} \\
& \\
\frac{< k_1 k_2 \rho_{\boldsymbol{k},1}>}{<k_2>} &
\frac{< (k_2^2 - k_2) \rho_{\boldsymbol{k},2}>}{<k_2>}
\end{array}
\right].
\]
By direct inspection, it is easy to see that the spectral radius
of $\boldsymbol{J}_s$ is 
equal to that of the matrix $\boldsymbol{J}_p$ defined in (\ref{eq:J_p});
this follows from the facts that $\boldsymbol{J}_s(i,i)=\boldsymbol{J}_p(i,i)$
for $i=1,2$ and 
$\boldsymbol{J}_s(1,2)\cdot\boldsymbol{J}_s(2,1)=\boldsymbol{J}_p(1,2)\cdot
\boldsymbol{J}_p(2,1)$.
Hence, 
as would be expected, we find that the recursive relations 
(\ref{eq:q_1})-(\ref{eq:q_2}) give the same global cascade condition 
(namely, $\sigma(\boldsymbol{J}_p)>1$) as the recursive relations 
(\ref{eq:g_1})-(\ref{eq:g_2}) obtained through utilizing generating functions. 
Nevertheless, the generating functions 
approach is useful in its own right as it enables 
quantifying the probability $P_{\textrm{trig}}$ 
of global cascades.

\begin{figure*}[!t]
\centering\subfigure[]{\hspace{-0.5cm} \includegraphics[totalheight=0.3\textheight,
width=.5\textwidth] {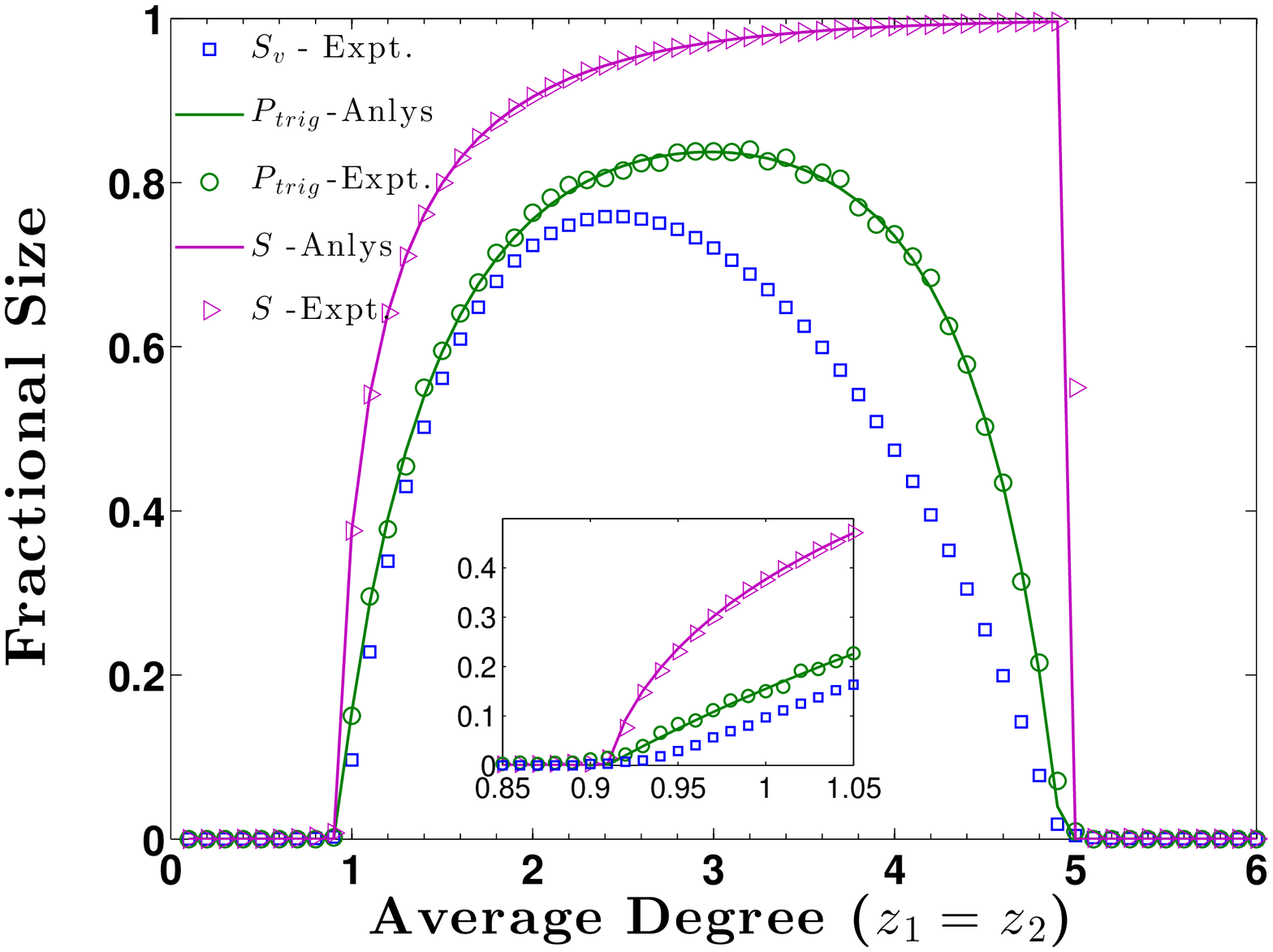} \label{fig:comp1}}
\subfigure[]{\hspace{-0.5cm}
\includegraphics[totalheight=0.3\textheight,
width=.5\textwidth] {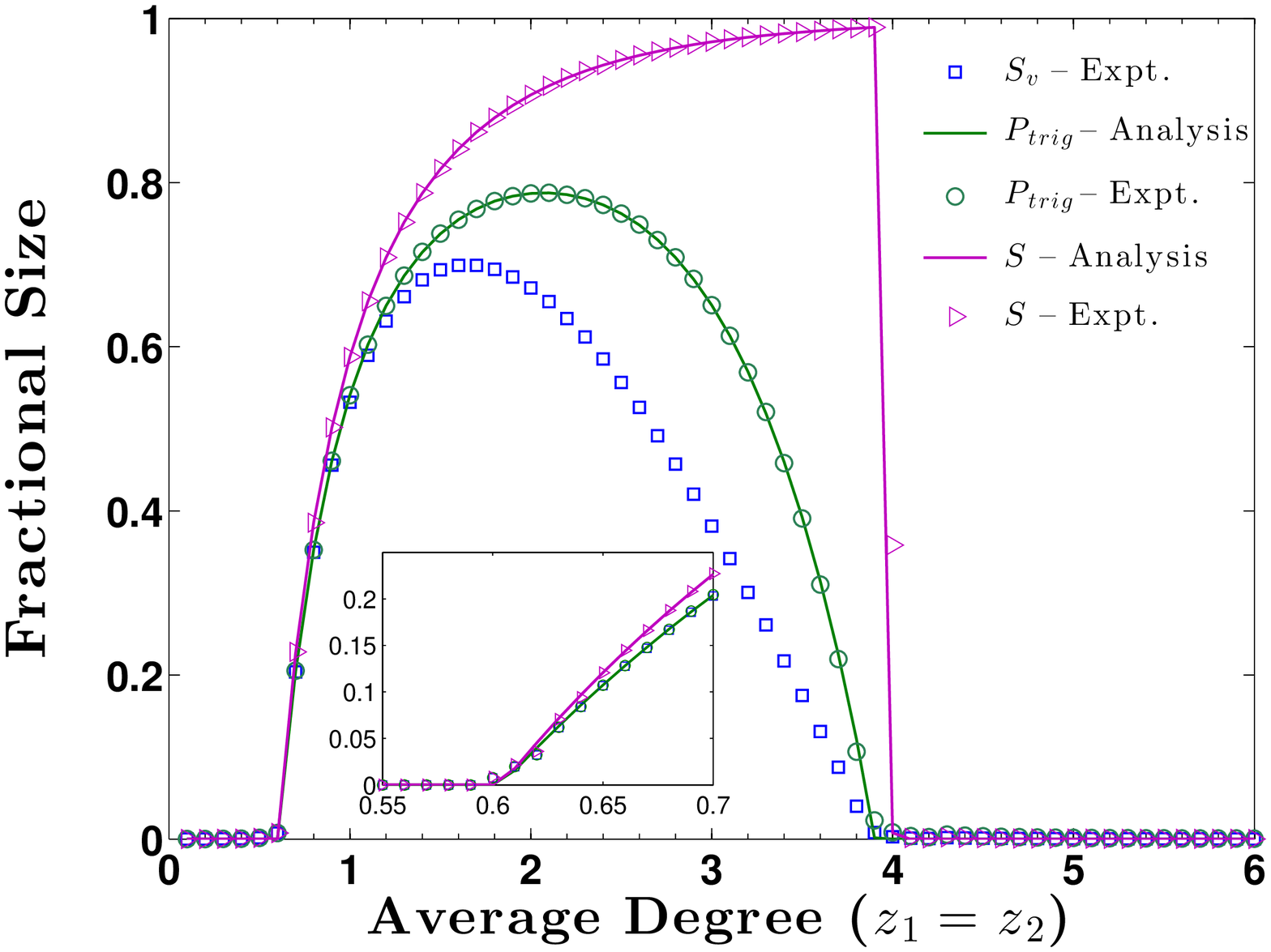}\label{fig:comp2}}
 \caption{(Color online) \sl The fractional 
size of the giant vulnerable cluster, $S_v$, the
final cascade size $S$ and the triggering probability
$P_{trig}$ are plotted for 
$n=5\times 10^5$, $\alpha=0.5$, $\tau$ is fixed at $\tau^{\star}=0.18$, and
$\mathbb{F}$ and
$\mathbb{W}$ are Erd\"{o}s-R\'enyi networks with mean 
degrees $z_1$ and $z_2$, respectively. The content parameter
is taken to be  
$(a)$ $c=0.25$ and $(b)$ $c=1$. 
The lines correspond to 
the analytical results obtained from Eqs. (\ref{eq:g_e(x)}) and (\ref{eq:S}), whereas symbols represent experimental results averaged 
over $100$ independent realizations 
for $S_v$ and $S$, and over $5,000$ realizations for $P_{trig}$. (Insets) The same plots 
at a higher resolution on $z_1=z_2$ near the  lower phase transition. 
}
\label{fig:comparison}
\end{figure*}

\section{Numerical Results}
\label{sec:Simu}

We now illustrate our findings by numerical simulations. In our first example we
consider $n=5 \times 10^5$ nodes in the physical network $\mathbb{W}$ and assume
that only half of these nodes are members of the network $\mathbb{F}$; i.e., we set 
$\alpha=0.5$. Following the references 
\cite{WattsExternal,Gleeson2008,BrummittLeeGoh,PayneDoddsEppstein} 
we fix the thresholds at $\tau=\tau^{\star}=0.18$ and assume that
both $\mathbb{F}$ and
$\mathbb{W}$ are Erd\"{o}s-R\'enyi networks \cite{Bollobas} with mean 
degrees $z_1$ and $z_2$, respectively. Figure \ref{fig:comparison} shows 
the fractional size of the giant vulnerable cluster $S_v$,  the triggering probability
$P_{trig}$, and the expected cascade size $S$ w.r.t. $z_1=z_2$ for two different contents. For the 
first content $\mathcal{C}_1$ we assume that 
 $c=0.25$, meaning that type-$2$ links
are four times as important as type-$1$ links in spreading this content,
and plot the corresponding results in Figure \ref{fig:comp1}. 
For the second content $\mathcal{C}_2$, we assume that both types of links are equivalent 
w.r.t. spreading the content, i.e., $c=1$, and
show the results in Figure \ref{fig:comp2}. In all cases,
lines correspond to our analysis results from Eqs. (\ref{eq:g_e(x)}) and (\ref{eq:S}),
whereas symbols are obtained from computer simulations: For each parameter set,
we generated independent realizations of the graphs $\mathbb{F}$ and $\mathbb{W}$, 
and then observed the cascade process over the graph $\mathbb{H}$ upon activating
an arbitrary node. The size $S_v$ of the giant vulnerable component is computed by 
finding the GSCC of the directed graph induced by the vulnerable nodes as 
described in Section \ref{sec:Results_cond_and_prob}.
The results are given by averaging over $100$ (resp. $5,000$) 
independent runs for $S_v$ and $S$ (resp. $P_{trig}$).

Figure \ref{fig:comparison} leads to a number of interesting  observations.
First, we see an excellent agreement between the analytical results and simulations,
confirming the validity of our analysis; the discrepancy near the upper phase
transition is due to finite size effect. Second, we see how content might impact 
the dynamics
of complex contagions over the same network. For content $\mathcal{C}_1$
we see that global cascades 
are possible when $1.0 \leq z_1=z_2 \leq 4.9$, whereas $\mathcal{C}_2$
can spread globally
only if  $0.7 \leq z_1=z_2 \leq 3.9$. We also see that on the range where
global cascades are possible for both contents (namely $1.0 \leq z_1=z_2 \leq 3.9$), the probability
of them taking place can still differ significantly; e.g., if $z_1=z_2=3$, we have $P_{trig}=0.84$ for 
$\mathcal{C}_1$, while for $\mathcal{C}_2$ we have $P_{trig}=0.65$.
Finally, simulations confirm the simultaneous appearance of the GSCC and GIN in the
subgraph of vulnerable nodes in $\mathbb{H}$ as understood from the identical 
parameter ranges that give positive values for $S$, $P_{trig}$ and $S_v$; see also the Inset of Figure \ref{fig:comp1}.
Therefore, the possibility of observing global cascades without a giant vulnerable cluster is
ruled out in our model, although this possibility exists in general. 

\begin{figure*}[!t]
\centering\subfigure[]{\hspace{-0.5cm} \includegraphics[totalheight=0.2\textheight,
width=.33\textwidth] {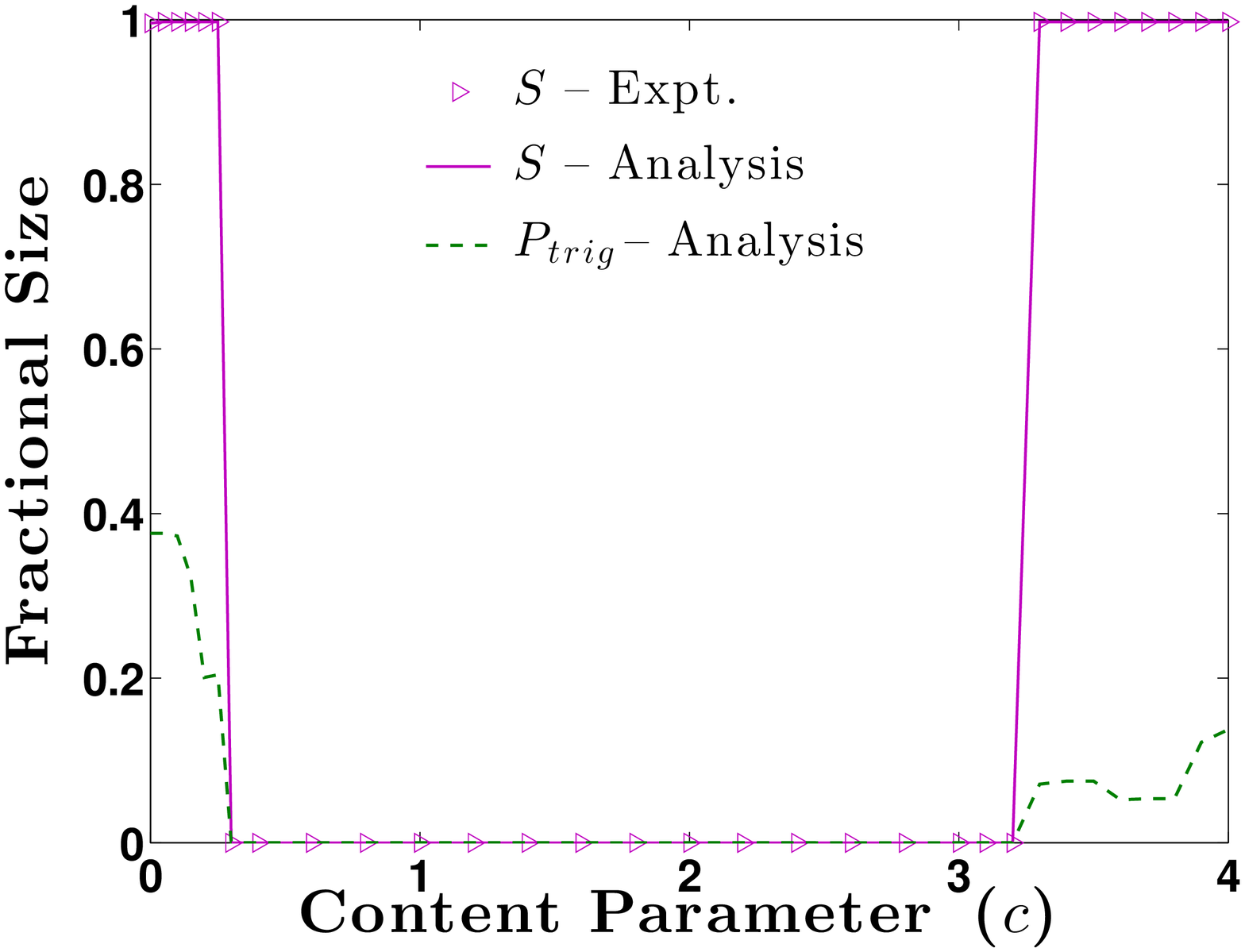} \label{fig:vary_c_1}} 
\subfigure[]{\hspace{-0.2cm} \includegraphics[totalheight=0.2\textheight,
width=.33\textwidth] {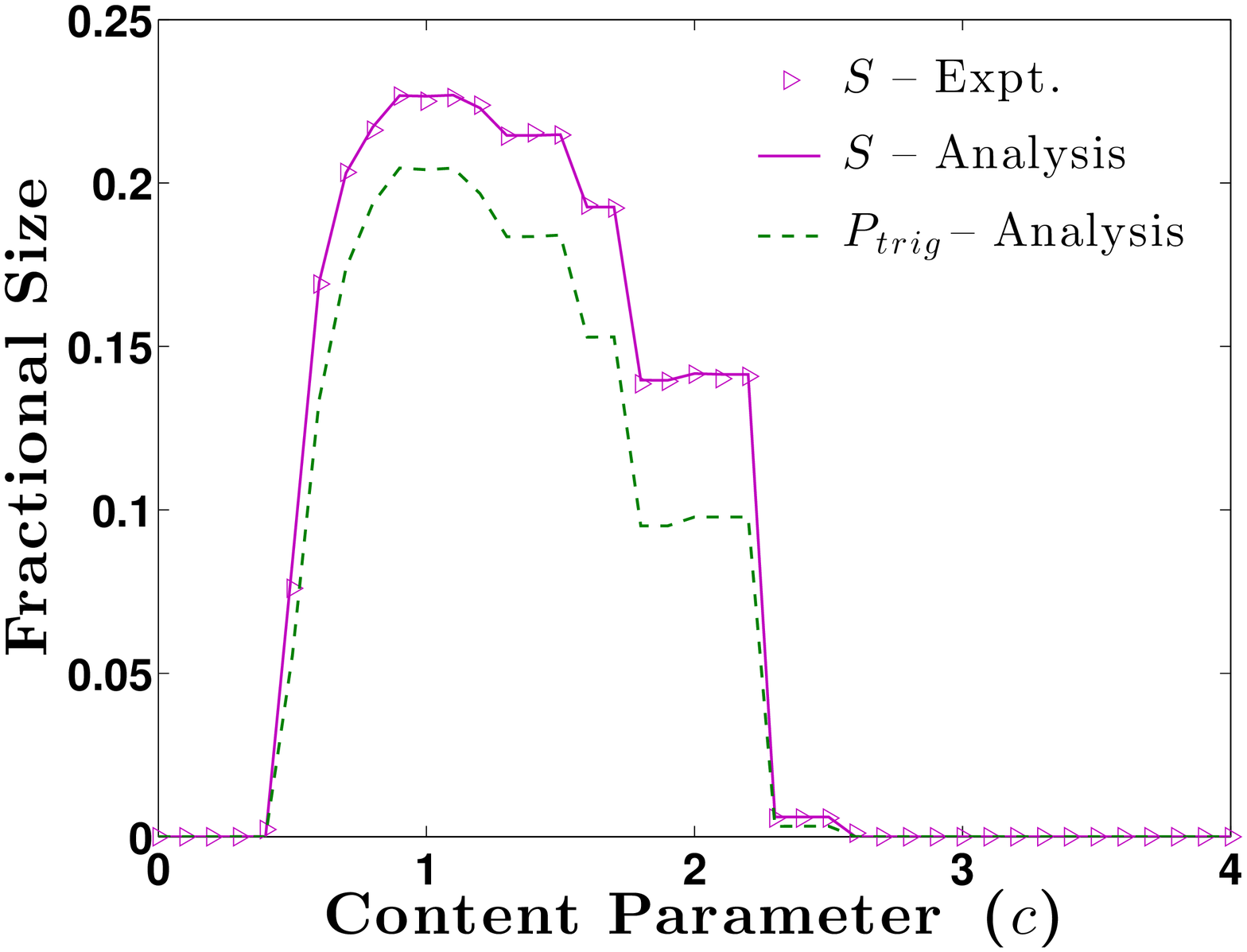} \label{fig:vary_c_2}}
\subfigure[]{\hspace{-0.2cm} \includegraphics[totalheight=0.2\textheight,
width=.33\textwidth] {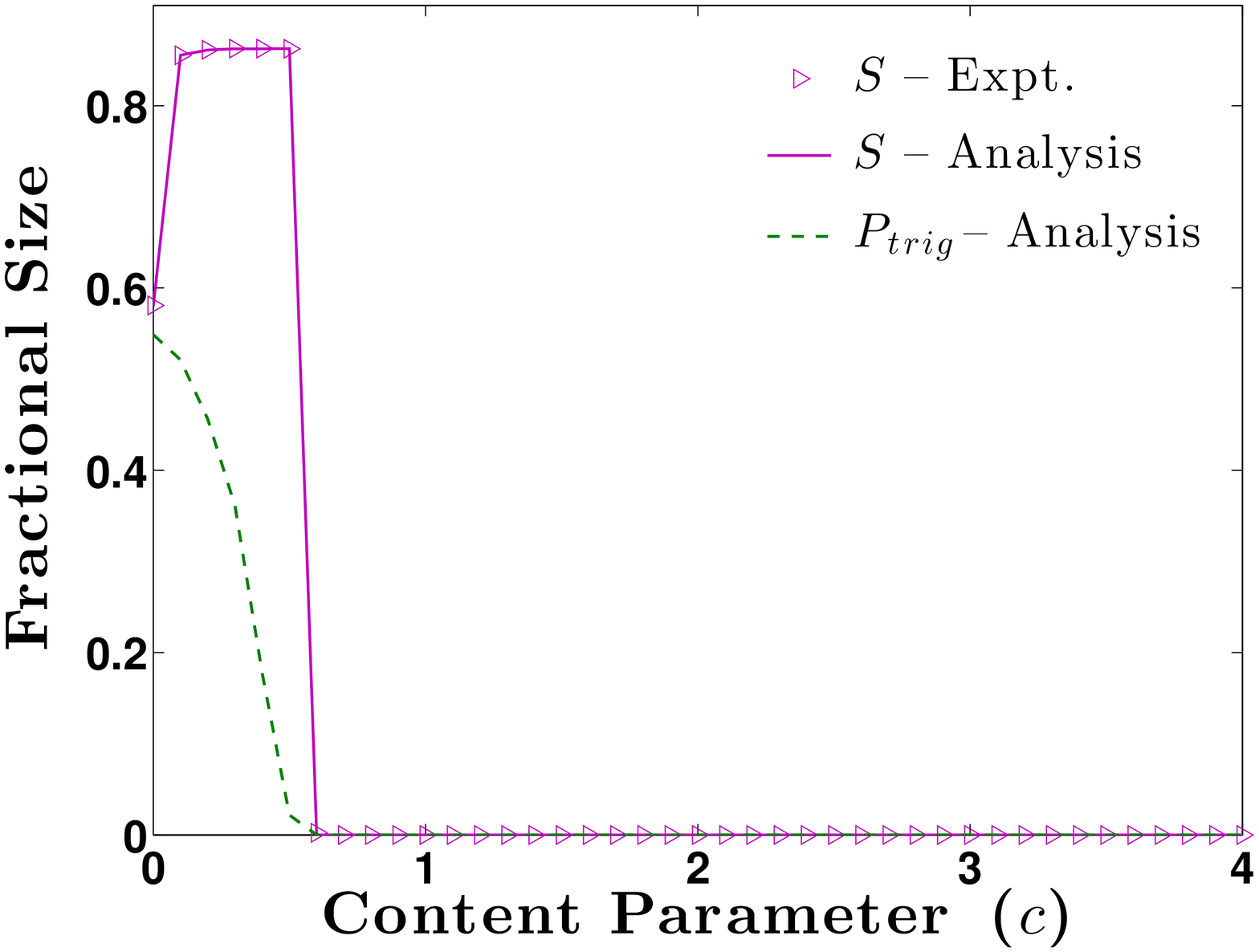} \label{fig:vary_c_3}}
\caption{(Color online) \sl We see the variation 
of cascade probability $P_{trig}$ and cascade size $S$ with respect
to the content parameter $c$, when $\alpha=0.5$, $\tau=\tau^{\star}=0.18$, and
$\mathbb{F}$ and $\mathbb{W}$ are Erd\"{o}s-R\'enyi networks with mean degrees
$z_1$ and $z_2$, respectively. 
We set $(a)$ $z_1=1.5$, $z_2=5.5$,  $(b)$  $z_1=z_2=0.7$, $(c)$ $z_1=6.0$, $z_2=1.5$. } 
\label{fig:varying_c}
\end{figure*}

For a better demonstration of the effect of content on 
the probability and size of global cascades, we now consider a
different experimental set-up. This time, for three different cases, we
fix all the parameters except the content-parameter $c$, and observe 
the variation of 
$P_{trig}$ and $S$ with respect to $c$. The results are depicted
in Figure \ref{fig:varying_c}. In all cases, we set $n=5\times 10^5$,
$\alpha=0.5$, $\tau =\tau^{\star}=0.18$, and assume that 
$\mathbb{F}$ and 
$\mathbb{W}$ are Erd\"{o}s-R\'enyi networks with mean degrees
$z_1$ and $z_2$, respectively. In Figure \ref{fig:vary_c_1}, we
consider the case $z_1=1.5$ and $z_2=5.5$, and see that
global cascades are possible only for $0 \leq c \leq 0.27$,
and $c \geq 3.22$, but no global cascade can take place in the range
$0.28 \leq c \leq 3.21$.  This can be explained as follows: 
When $c$ is too small, the spreading of the
content is governed solely by network $\mathbb{W}$ (with average 
degree $z_2= 5.5$), on which large global cascades are possible
with very low probability; in other words network $\mathbb{W}$ 
is close to the upper phase transition threshold. 
As $c$ gets larger, the effect of $\mathbb{F}-$links 
becomes considerable, and the connectivity of the overall 
network (with respect to the spread of the current content) increases. 
This eventually causes global cascades to 
disappear due to the high local stability (i.e., connectivity) of the nodes. 
However, further increase in $c$ shifts the bias towards $\mathbb{F}-$links,
and due to lower average degree in $\mathbb{F}$, this causes a decrease
in the local stability of the nodes and brings global cascades back 
to the existence. 
 
In Figure \ref{fig:vary_c_2}, we set $z_1=z_2=0.7$. This time, we see
an exactly opposite dependence of cascade sizes on the parameter $c$.
Namely, global cascades are not possible when $c$ is too small or too large, but 
they do take place in the interval $0.5 \leq c \leq 2.5$. This is because,
under the current setting, both networks
$\mathbb{F}$ and $\mathbb{W}$ have limited connectivity, so that global cascades
do not take place in either of the networks separately. Therefore, if $c$
is too small (resp. too large), only $\mathbb{W}$  (resp. $\mathbb{F}$)
can spread the content and  all triggering 
events have finite size as confirmed in the plot by the non-existence of global cascades
for $c \leq 0.4$ and $c \geq 2.6$. But, for $c$ relatively close to unity, the
two networks spread the content collaboratively,
yielding a high enough connectivity in the overall network $\mathbb{H}$ 
to achieve a positive
probability of global cascades.

The situation is somewhat different in the case of Figure \ref{fig:vary_c_3}, where
we have $z_1=6.0$ and $z_2=1.5$: For contents
that mainly spread over $\mathbb{W}$-links, i.e., for $c$ close to zero, 
global spreading events  take place with positive probability since network 
$\mathbb{W}$ (with average degree $z_2=1.5$) satisfies the global spreading
condition. However, as $c$ gets larger, the high average degree in 
network $\mathbb{F}$ (and, thus the high 
local stability of the nodes) makes it harder for the content
to spread in the overall network $\mathbb{H}$, eventually causing
the probability of global cascades drop to zero. This is confirmed 
in Figure \ref{fig:vary_c_3} as we see that global cascades 
take place only for contents with $c \leq 0.5$ and any content
with $c \geq 0.6$ dies-out before reaching a non-trivial fraction of
the network.

We continue our simulation study by depicting 
the variation of the {\em cascade window} with respect to content 
parameter $c$ in Figure \ref{fig:window}. 
We see that for each $\tau^{\star}$, the parameter $c$ changes the range of $z_1=z_2$
for which global cascades can occur in a {\em non-trivial} way. For instance, none of the 
three regions cover one another. In fact, the
cascade window for $c=4$ is contained in that of $c=1$ for most of the $\tau^{\star}$ values,
but, with $z_1=z_2=4$ and $\tau^{\star}$ in 
$(0.17, 0.18)$ cascades do not occur for $c=1$ while they do for $c=4$. We also see that the content 
parameter $c$ can effect the maximum threshold $\tau^{\star}$ for which global cascades are possible.
When $c=1$ and $c=0.1$, cascades can take place for $\tau^{\star} \leq 0.25$, whereas the
upper-bound is reduced to $0.23$ for $c=4$.

Finally, we test our theory for networks which are {\em not} locally tree like. In fact, most 
{\em real} networks are known \cite{SerranoBoguna} to exhibit a phenomenon often
called {\em clustering} (or transitivity), informally defined as the propensity of a node's
neighbors to be neighbors as well. Since our theory is developed for networks that do not
have clustering, we do not expect our results to provide good estimations for clustered networks;
in the case of Watts' threshold model, it is already shown \cite{HackettMelnikGleeson} 
that clustering can have a significant impact on the size of global cascades. 
Nevertheless, we would like to provide the first step in showing the effect of clustering
on content-dependent cascading processes in multiplex networks. 

To this end, we generate random {\em clustered} networks 
$\mathbb{F}$ and $\mathbb{W}$ as prescribed by Newman
\cite{NewmanCluster} and Miller \cite{Miller}. Namely, we consider
distributions $p_{st}^f$ and $p_{st}^w$
that give the probability of a node being connected to $s$ single edges and $t$ triangles;
conventional degree distributions are then given by $p_k^f = \sum_{s,t} p_{st}^f \delta_{k,s+2t}$
and $p_k^w = \sum_{s,t} p_{st}^w \delta_{k,s+2t}$.
For convenience, we consider a doubly Poisson distribution for $p_{st}$; namely, we set
\begin{equation}
p_{st}^f = e^{-z_1} \frac{z_1^s}{s!} \cdot e^{-z_1/2} \frac{(z_1/2)^t}{t!}, \quad s,t = 0,1, \ldots
\label{eq:deg_p_st}
\end{equation}
and define $p_{st}^w$ similarly with $z_1$ replaced by $z_2$. Notice that average
degrees are now given by $2 z_1$ and $2 z_2$ in networks $\mathbb{F}$ and 
$\mathbb{W}$, respectively. With $n=10^5$, $\alpha=0.5$, $\tau=\tau^{\star}=0.18$
and $c=0.5$, we show in Figure \ref{fig:cluster} the variation of the cascade size
$S$ with respect to average degrees $2z_1=2z_2$. As before, the line corresponds 
to the analytical solution (obtained from Eq. (\ref{eq:S})),
whereas symbols are obtained from simulations by averaging over
$50$ experiments for each point. Also, in the Inset of Figure \ref{fig:cluster}, we plot
the average clustering coefficient \cite{NewmanCluster} observed 
for networks $\mathbb{W}$ and $\mathbb{F}$; with $z_1=z_2$ both networks have
(statistically) identical
clustering coefficients. 

As expected, we do not see a good match between the predictions of our analysis
(zero-clustering) and the actual cascade size from experiments (positive clustering). 
However, these results
agree with the double faceted picture drawn in \cite{HackettMelnikGleeson} 
for the effect of clustering on cascade sizes in Watts' model: When average degrees
are small, clustering {\em decreases} the expected size of global cascades, whereas 
after a certain value of average degree, 
clustering {\em increases} the expected cascade size.

\begin{figure*}[!t]
\centering\subfigure[]{\hspace{-0.5cm} \includegraphics[totalheight=0.3\textheight,
width=.5\textwidth] {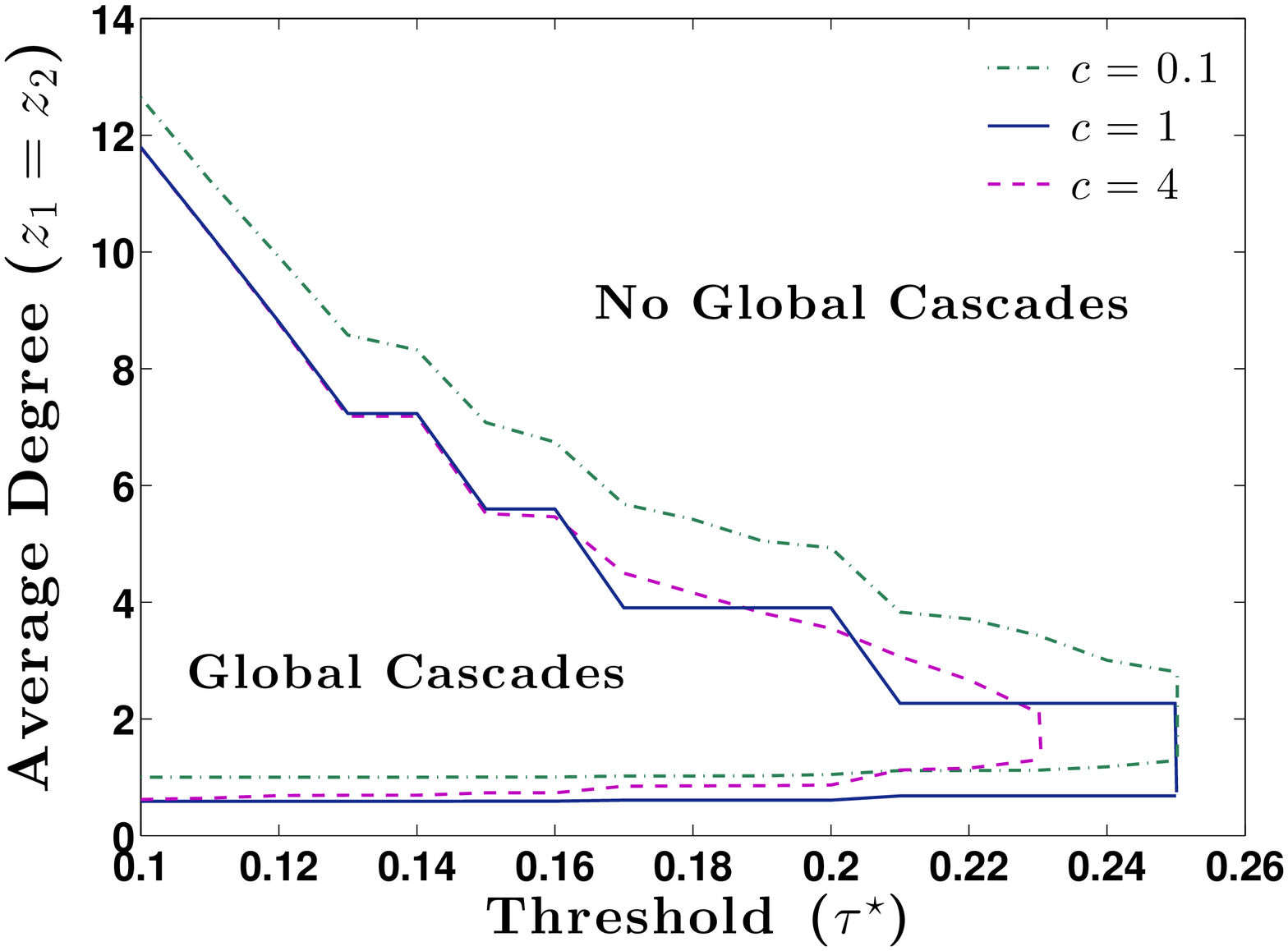} \label{fig:window}} 
\subfigure[]{\hspace{-0.2cm} \includegraphics[totalheight=0.3\textheight,
width=.5\textwidth] {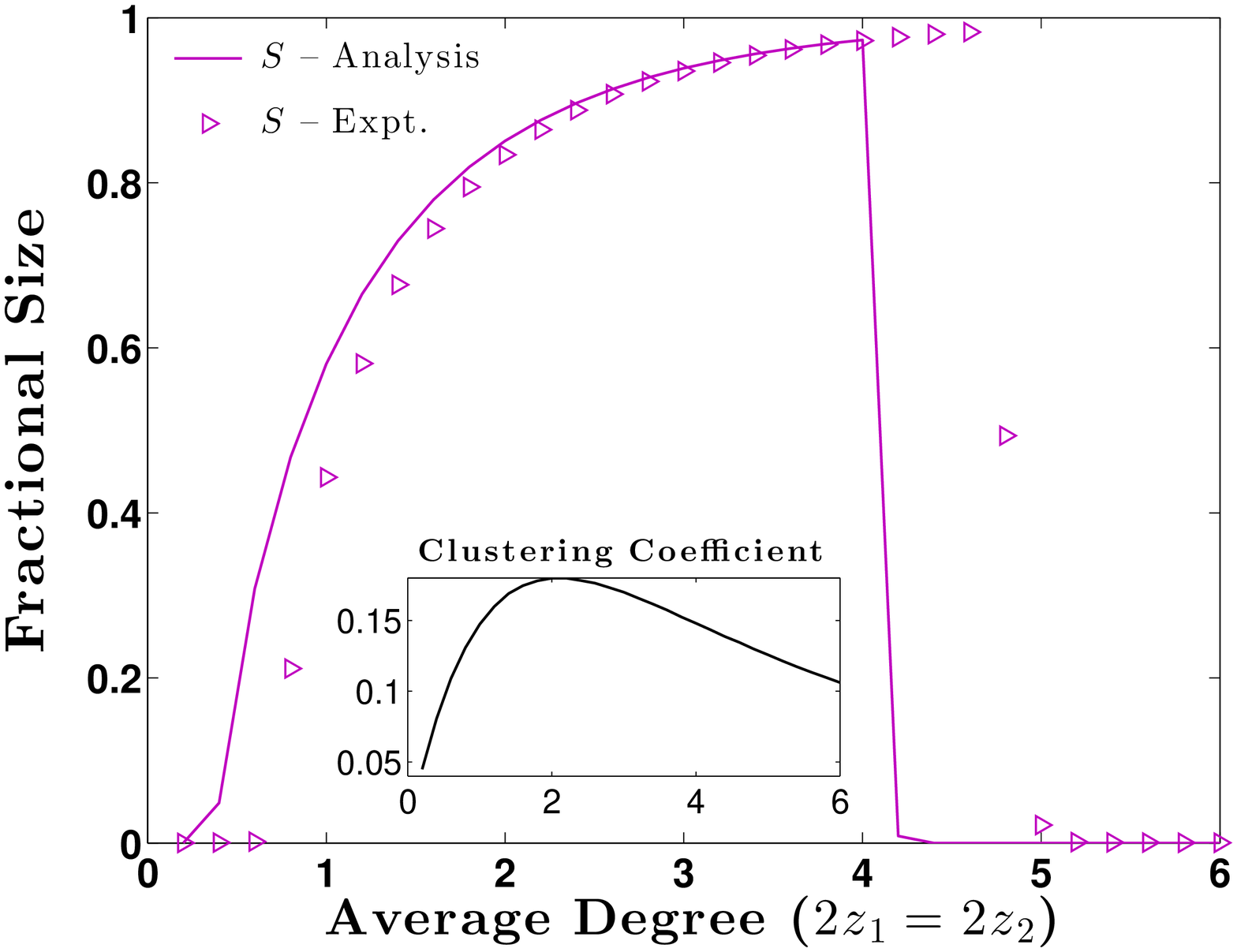} \label{fig:cluster}}
\caption{(Color online) \sl (a) We see the {\em cascade windows} for several $c$
values when 
$\alpha=0.5$, $\tau = \tau ^ {\star}$, and 
$\mathbb{F}$ and
$\mathbb{W}$ are Erd\"{o}s-R\'enyi networks with mean 
degrees $z_1$ and $z_2$, respectively. In other words,
the lines enclose the 
region of the $(\tau^{\star}, z_1=z_2)$ plane for which the global cascade condition
$\sigma(\boldsymbol{J}_p) >1$ is satisfied; outside the boundary, we have
$\sigma(\boldsymbol{J}_p) \leq 1$ and global cascades can not take place. 
(b) Global cascade size $S$ when $\mathbb{F}$ and $\mathbb{W}$ are 
random clustered networks with degree distributions given by (\ref{eq:deg_p_st}).
We take $n=10^5$, $\alpha=0.5$, $\tau=\tau^{\star}=0.18$ and $c=0.5$. The line corresponds
to analytical prediction from (\ref{eq:S}), whereas symbols are obtained from simulations 
by averaging over 50 independent realizations. (Inset) Average clustering
coefficient \cite{NewmanCluster} versus $2z_1=2z_2$.}
\label{fig:CascadeWindow_and_Cluster}
\end{figure*}
  
\section{Conclusion}
\label{sec:Conclusion}

We have determined the condition, probability and the size
of global cascades in random networks with classified links.
This is done
under a new contagion model
where nodes
switch state when their
{\em perceived} (content-dependent) proportion of active neighbors 
exceed a
certain threshold.  
Our results highlight the effect of content and link classification 
in the characteristics of global cascades, and show how different 
content may have different spreading characteristics
over the same network. Also, the results given here 
extend the existing work on complex contagions to 
multiple overlay networks whose vertices are not disjoint.

Our findings also contain some of the
existing results as special cases. For instance, our results 
may be applied to a wide range of processes on the network $\mathbb{H}$ by
appropriately selecting the neighborhood response function $F(\boldsymbol{m},\boldsymbol{k})$. 
In particular, the 
general results of Section \ref{sec:Results} include the solutions for
bond percolation, and {\em simple} contagion \footnote{Simple contagions are
defined as diffusion processes where nodes become infected (or active) after only 
one incident of contact with an infected neighbor. Examples 
include spread of diseases and information.}  processes by setting 
$\rho_{\boldsymbol{k},1}=\rho_{\boldsymbol{k},2}=T$ for some transmissibility
$T$ in $[0, 1]$.
The threshold model of Watts \cite{WattsExternal}
is also covered by our theory by setting the content parameter
$c$ to unity in all cases. 

We believe that the results presented here give some interesting insights 
into the cascade processes in complex networks. In particular, our results might
help better understand such processes and may lead to more efficient 
control of them. Controlling cascade processes is particularly 
relevant when dealing with cascading failures in interdependent structures
as well as when marketing a certain consumer product. 
Finally, the formulation presented here opens many new questions in the field.
For instance, the dynamics of cascade processes are yet to be investigated on
clustered or degree-correlated networks under the content-dependent threshold
model introduced here. 
Another challenging problem would be to formalize 
the results given in this paper {\em without} using a mean-field approach;
in fact, very recently Lelarge and co-workers 
\cite{Lelarge,CoupechouxLelarge} have obtained rigorous results for the condition
and size of global cascades in Watts' threshold model.

\section*{Acknowledgments}
This research was supported in part by CyLab at Carnegie Mellon
under grant DAAD19-02-1-0389 from the US Army Research Office.
The views and conclusions contained in this document are those of the authors
and should not be interpreted as representing the official policies,
either expressed or implied, of any sponsoring institution, the
U.S. government or any other entity.

\bibliography{references}

\end{document}